  \documentstyle[preprint,aps]{revtex}
  
\begin{document}
  \draft
  \preprint{\begin{tabular}{l}YUMS 93-15\\SNUTP 93-44\\(July
1993)\end{tabular}}
  \vskip 2.5cm
  \title{\bf Probing $WW\gamma$ Couplings from
  $\sigma_{tot}(W)/\sigma_{tot}(Z)$\\
  in High Energy $ep$ Collisions\footnote{Talk given by C.S. Kim at the 2nd
  International Workshop on future $e^+ e^-$ colliders at Waikoloa, Hawaii
  in April 1993}}
  \author{C. S. Kim$^{1}$, Jungil Lee$^{2}$, and H. S. Song$^{2}$}
  \address{
  $^{1}$ Department of Physics, Yonsei University, Seoul 120-749, Korea\\
  $^{2}$ Dept. of Physics, Seoul National University, Seoul 151-742, Korea}
  \maketitle
  \vskip 4cm
  \begin{abstract}

  We investigate the sensitivity of total cross sections of
  $e+p \rightarrow W,Z$ to CP-conserving non-standard $WW\gamma$
  couplings. We include all the important production mechanisms and
  study the dependence of the total $W$ cross sections on the
  anomalous $WW\gamma$ couplings, $\kappa$ and $\lambda$.
  We argue that the ratio of $W$ and $Z$ production cross sections is
  particularly well suited, being relatively insensitive to uncertainties
  in the theoretical and experimental parameters.
  \end{abstract}
  \renewcommand{\thepage}{}
  \newpage
  \renewcommand{\thepage}{\arabic{page}}\setcounter{page}{1}

  In order to parametrize non-standard $WW\gamma$ couplings, it is important
  to know what sort of additional couplings
  can arise once the restrictions due to gauge
  invariance are lifted. As has been previously shown$^1$,
  there can be $14$ or more non-standard couplings in the most general
  case. To keep the analysis manageable, we restrict ourselves to CP
  and $U(1)$ conserving couplings. This restriction leads to just two
  anomalous form-factors traditionally denoted by $\lambda$ and $\kappa$,
  which can be related$^2$ to the anomalous
  electric quadrupole and the anomalous magnetic dipole moment
  of the $W$. In the Standard Model (SM) at tree level
  $\lambda=0$ and $\kappa=1$. At present the best experimental limits
  $-3.6<\lambda<3.5$ and $-3.5<\kappa<5.9$ are from a recent analysis
  by UA(2) collaboration$^3$.
  The total production of $W$ boson at $ep$ colliders will
  provide a very precise test of the structure of SM triple-boson
  $WW\gamma$ vertex.
  A measurement of the anomalous coupling
  in the $WW\gamma$ vertex at $ep$ colliders can best be achieved by
  considering the ratio of the $W$ and $Z$ production cross sections.
  The advantage of using a cross section ratio is that uncertainties
  from the luminosity, structure functions, higher-order corrections,
  QCD scale, etc. tend to cancel$^4$.
  We include both the lowest-order resolved
  processes and the dominant direct photoproduction processes. Care
  must be taken to avoid double counting those phase-space regions
  of the direct processes which are already included in the resolved
  processes.

  We first focus on the total production of $W$ and $Z$ in $ep$ collisions.
  In the short term these processes will be studied at
  HERA($E_{e}=30~$GeV$, E_{p}=820~$GeV$, {\cal L}=200~$pb$^{-1}~$yr$^{-1}$),
  while in the long term availability of
  LEP~$\times$~LHC(
  $E_{e}=50~$GeV$, E_{p}=8000~$GeV$, {\cal L}=1000~$pb$^{-1}~$yr$^{-1}$)
  collider will give collision energies in excess of 1~TeV.
  We calculate the total cross sections for the five different processes
  which contribute to single $W$ and $Z$ production at $ep$ colliders.
  From the sum of these contributions we then calculate the ratio
  $\sigma(W)/\sigma(Z)$ as a function of the anomalous
  $WW\gamma$ coupling parameters $\kappa$ and $\lambda$.
  The processes are
  \renewcommand{\theequation}{1\alph{equation}}\setcounter{equation}{0}
  \begin{eqnarray}
  e^{-}+p &\rightarrow& e^{-}+W^{\pm}+X,\\
  &\rightarrow& \nu+W^{-}+X,\\
  &\rightarrow& e^{-}+Z+X ~(Z\mbox{ from hadronic vertex}),\\
  \label{eq:Zh}
  &\rightarrow& e^{-}+Z+X ~(Z\mbox{ from leptonic vertex}),\\
  \label{eq:Zl}
  &\rightarrow& \nu+Z+X.
  \end{eqnarray}

  The largest contributions for $W$ and $Z$ productions come from
  the processes (1a) and (1c) which are dominated by
  the real photon exchange Feynman diagrams with a photon emitted from
  the incoming electron,
  $e^{-}+p \rightarrow \gamma_{/e}+p \rightarrow V+X$.
  The dominant subprocesses for
  $\gamma+p \rightarrow V+X$ would appear to be the lowest order
  $\bar{q}^{(\prime)}_{/\gamma}+q \rightarrow V$, where $q_{/\gamma}$
  is a quark inside the photon. However this may not be strictly true,
  even at very high energies, since quarks inside the photon $q_{/\gamma}$
  exist mainly through the evolution $\gamma \rightarrow q\bar{q}$.
  Hence the direct process
  $\gamma+q \rightarrow q^{(\prime)}+V$
  could be competitive with the lowest order contribution
  $\bar{q}^{(\prime)}+q \rightarrow V$.
  This raises the subtle question of double counting$^4$.
  Certain kinematic regions of the direct processes contribute to the
  evolution of $q_{/\gamma}$ which is already included in the lowest
  order process. Both double counting and the mass singularities are
  removed$^5$ if we subtract the contribution of
  $\gamma+q \rightarrow q^{(\prime)}+V$
  in which the $\hat{t}$-channel-exchanged quark is on-shell
  and collinear with the parent photon. Thus the singularity subtracted
  lowest order contribution from the subprocesses
  $\bar{q}^{(\prime)}_{/\gamma}+q \rightarrow V$ is
  \renewcommand{\theequation}{\arabic{equation}}\setcounter{equation}{1}
  \begin{equation}
  \sigma^{L}(e^{-}+~p \rightarrow V+X)
  =\frac{{\cal C}^{L}_{V}}{s}\int_{m^{2}_{V}/s}^{1}
  \frac{dx_{1}}{x_{1}}
  \left[\sum_{qq^{\prime}}
  (f_{q_{/e}}-\tilde{f}_{q_{/e}})(x_1)
  f_{q^\prime_{/p}}(\frac{m^{2}_{V}}{x_{1}s})
  +(q \leftrightarrow q^\prime)\right],
  \end{equation}
  where ${\cal C}^{L}_{V=W,Z}$ are defined in Ref. 4.
  The electron structure functions $f_{q/e}$ (and $\tilde{f}_{q/e}$)
  are obtained as usual by convoluting the photon structure functions
  $f_{q_{/\gamma}}$ with the
  Weiz\"{a}cker-Williams approximation of (quasi-real)
  photon radiation.
  The part of photon structure function,
  $\tilde{f}_{q_{/\gamma}}$, results from photon splitting at large
  $x$ (with large momentum transfer)$^4$.
  To obtain the total cross section of processes (1a) and (1c), we add to the
  cross section in (2) the contribution from the direct subprocesses,
  $\gamma+q \rightarrow q^{(\prime)}+V$,
  \renewcommand{\theequation}{3\alph{equation}}\setcounter{equation}{0}
  \begin{equation}
  \sigma^{D}(e^{-}+~p \rightarrow V+X)
  =\frac{{\cal C}^{D}_{V}}{s}
  \int_{m^{2}_{V}/s}^{1}\frac{dx_{1}}{x_{1}}
  \int_{m^{2}_{V}/x_{1}s}^{1}\frac{dx_{2}}{x_{2}}
  \left[\sum_{q}
  f_{\gamma_{/e}}(x_1)f_{q_{/p}}(x_{2})
  \right]
  \eta_{V}(\hat{s}),
  \end{equation}
  where the integrated hard scattering cross sections $\eta_V$ are
  \begin{eqnarray}
  &\eta&_{V=Z}(\hat{s},m^{2}_{Z},\Lambda^{2})
  =(1-2z+2z^{2})\log\left(\frac{\hat{s}-m^{2}_{Z}}{\Lambda^{2}}\right)
  +\frac{1}{2}(1+2z-3z^{2}),\nonumber\\
  &\eta&_{V=W}(\hat{s},m^{2}_{W},\Lambda^{2},Q=|e_{q}|,\kappa,\lambda)
  =(Q-1)^{2}(1-2z+2z^{2})
  \log\left(\frac{\hat{s}-m^{2}_{W}}{\Lambda^{2}}\right)\\
  &&\hspace{.15in}
  -\left[(1-2z+2z^{2})-2Q(1+\kappa+2z^{2})
  +\frac{(1-\kappa)^{2}}{4z}-\frac{(1+\kappa)^{2}}{4}
  \right]\log{z}\nonumber\\
  &&\hspace{.15in}
  +\left[\left(2\kappa+\frac{(1-\kappa)^{2}}{16}\right)\frac{1}{z}
  +\left(\frac{1}{2}+\frac{3(1+Q^{2})}{2}\right)z
  +(1+\kappa)Q-\frac{(1-\kappa)^{2}}{16}+\frac{Q^{2}}{2}
  \right](1-z)\nonumber\\
  &&\hspace{.15in}
  -\frac{\lambda^{2}}{4z^{2}}(z^{2}-2z\log{z}-1)
  +\frac{\lambda}{16z}(2\kappa+\lambda-2)
  \left[(z-1)(z-9)+4(z+1)\log{z}\right],\nonumber
  \end{eqnarray}
  and ${\cal C}^{D}_{V=W,Z}$ are defined in Ref. 4.

  The processes (1b) and (1d) are dominated by configurations
  where a~(quasi-real) photon is emitted~(either elastically or
  quasi-elastically) from the incoming proton and subsequently scatters
  off the incoming electron, i.e.
  $e^{-}+p \rightarrow e^{-}+\gamma_{/p} \rightarrow e^{-}~(~$or$~\nu)+V.$
  For the elastic photon, the cross section can be computed using the
  electrical and magnetic form factors of the proton.
  For the quasi-elastic scattering photon, the experimental information$^6$
  on electromagnetic structure functions $W_{1}$ and
  $W_{2}$ can be used, following Ref.~7. The hard scattering
  cross section for $e^{-}+\gamma_{/p} \rightarrow e^{-}~(~$or$~\nu)+V$
  is given by
  \renewcommand{\theequation}{\arabic{equation}}\setcounter{equation}{3}
  \begin{equation}
  \hat{\sigma}(e^{-}+\gamma_{/p} \rightarrow e^{-}~(~\mbox{or}~\nu)+V)
  =\frac{{\cal C}^{D}_{V}}{\hat{s}}
  \eta_{V}(Q=|e_{q}|=1).
  \end{equation}
  Finally for process (1e), which is a pure charged current process, we simply
  use the results of Bauer {\it et. al.}$^7$ to add to the
  contributions from (1c) and (1d). The contribution from this process
  to the total $Z$ production cross section is almost negligible even at
  LEP~$\times$~LHC $ep$ collider energies.
  With the anticipated luminocities
  the total $Z$ production cross section corresponds to
  84~events/yr~(HERA) and 5400~events/yr~(LEP~$\times$~LHC).
  After including a $6.7~\%$ leptonic branching
  ratio~(i.e.~$Z \rightarrow e^{+}e^{-},\mu^{+}\mu^{-}$),
  the event numbers become about 6~events/yr(HERA) and
  360~events/yr(LEP~$\times$~LHC).

  In Fig. 1, we show the ratio of $\sigma(W^{\pm})/\sigma(Z)$ and
  $\sigma(W^{-})/\sigma(Z)$ as a function of $\kappa$ and $\lambda$.
  Rather than vary both parameters simultaneously, we first set
  $\kappa$ to its Standard Model value and then vary $\lambda$ and
  vice versa. The error range represents
  the variation in the cross section by varying the theoretical input
  parameters as follows~:~$m^{2}_{V}/10\leq{Q}^{2}\leq{m}^{2}_{V}$, photon
  structure functions $f_{q_{/\gamma}}$ from DG$^8$ and
  DO$+$VMD$^9$, and proton structure functions $f_{q_{/p}}$
  from EHLQ1$^{10}$ and HMRS(B)$^{11}$.
  It is important to note that
  once photoproduction experiments at
  HERA determine $f_{q_{/p}}$ and $f_{q_{/\gamma}}$ more precisely,
  we will be able to predict the total cross sections for each process
  with much greater accuracy.
  After 5 years of running, HERA will produce about 30
  $e+p \rightarrow Z+X \rightarrow l^{+}+l^{-}+X$ events,
  and this will enable us to determine$^{12}$ $\kappa$ and $\lambda$
  with a precision of order
  $\Delta\kappa\approx{\pm}0.3~\mbox{for}\hspace{.1in}\lambda=0,
  ~\Delta\lambda\approx{\pm}0.8~\mbox{for}\hspace{.1in}\kappa=1$.
  At LEP~$\times$~LHC, one year's running will give
  $\Delta\kappa\approx{\pm}0.2~\mbox{for}\hspace{.1in}\lambda=0,
  ~\Delta\lambda\approx{\pm}0.3~\mbox{for}\hspace{.1in}\kappa=1$.

  \vskip 2.5cm
  \noindent{\bf Acknowledgements}

  The work was supported in part by the Korea Science and Engineering
  Foundation and in part by the Korean Ministry of Education.
  The work of CSK was also supported in part by the Center
  for Theoretical Physics at Seoul National University and in part by a
  Yonsei University Faculty Research Grant.

  \newpage

  \noindent{\bf References}
  \vskip 1cm
  \noindent 1. K. Hagiwara {\it et. al.}, Nucl.\ Phys.\ {\bf B282} (1987) 253.

  \noindent 2. H. Aronson, Phys.\ Rev.\ {\bf 186} (1969) 1434;
  K. J. Kim and Y. S. Tsai, Phys.\ Rev.\ {\bf D7} (1973) 3710.

  \noindent 3. J. Alitti {\it et. al.}, Phys.\ Lett. {\bf B277} (1992) 194.

  \noindent 4. C. S. Kim and W. J. Stirling, Z. Phys. {\bf C53} (1992) 601.

  \noindent 5. F. I. Olness and W.-K. Tung, Nucl.\ Phys. {\bf B308} (1988) 813.

  \noindent 6. S. Stein {\it et. al.}, Phys.\ Rev. {\bf D12} (1975) 1884;
  A. Bodek {\it et. al.}, Phys. Rev. {\bf D20} (1971) 1471.

  \noindent 7. U. Bauer, J. A. M. Vermaseren and D. Zeppenfeld, Nucl.\ Phys.
  {\bf B375} (1972) 3.

  \noindent 8. M. Drees and Grassie, Z.\ Phys. {\bf C28} (1985) 451.

  \noindent 9. D. W. Duke and J. F. Owens, Phys.\ Rev. {\bf D26} (1982) 1600.

  \noindent 10. E. Eichten, I. Hinchliffe, K. D. Lane and Quigg,
  Rev.\ Mod.\ Phys.{\bf 56} (1984) 579.

  \noindent 11. P. N. Harriman, A. D. Martin, W. J. Stirling
  and R. G. Roberts,Phys.\ Rev. {\bf D42} (1990) 798.

  \noindent 12. For more details, see C.S. Kim, Jeongil Lee and H.S. Song,
  YUMS 93--14 (July 1993)

  \newpage
  \noindent{\bf Figure Captions}
  \vskip 1cm

  Fig. 1. Total cross section ratios of $\sigma(W^{\pm})/\sigma(Z)$ and
  $\sigma(W^{-})/\sigma(Z)$ as a function of (a) $\kappa$, and (b) $\lambda$
  at the HERA and LEP$\times$LHC $ep$ colliders.
  Rather than vary both parameters simultaneously, we first set
  $\kappa$ to its Standard Model value and then vary $\lambda$ and
  vice versa. The error range represents
  the variation in the cross section by varying the theoretical input
  parameters as follows~:~$m^{2}_{V}/10\leq{Q}^{2}\leq{m}^{2}_{V}$, photon
  structure functions $f_{q_{/\gamma}}$ from DG$^8$ and
  DO$+$VMD$^9$, and proton structure functions $f_{q_{/p}}$
  from EHLQ1$^{10}$ and HMRS(B)$^{11}$.

  \end{document}